# BioSFQ Circuit Family for Neuromorphic Computing: Bridging Digital and Analog Domains of Superconductor Technologies

Vasili K. Semenov, Evan B. Golden, and Sergey K. Tolpygo, *Senior Member, IEEE*

*Abstract*—Superconductor single flux quantum (SFQ) technology is attractive for neuromorphic computing due to low energy dissipation and high, potentially up to 100 GHz, clock rates. We have recently suggested a new family of bioSFQ circuits (V.K. Semenov *et al.*, *IEEE TAS*, vol. 32, no. 4, 1400105, 2022) where information is stored as a value of current in a superconducting loop and transferred as a rate of SFQ pulses propagating between the loops. This approach, in the simplest case dealing with positive numbers, requires single-line transfer channels. In the more general case of bipolar numbers, it requires dual-rail transfer channels. To address this need, we have developed a new comparator with a dual-rail output. This comparator is an essential part of a bipolar multiplier that has been designed, fabricated, and tested. We discuss bioSFQ circuits for implementing an analog bipolar divide operation $Y/X$ and a square root operation $\sqrt{X}$. We discuss strategic advantages of the suggested bioSFQ approach, e.g., an inherently asynchronous character of bioSFQ cells which do not require explicit clock signals. As a result, bioSFQ circuits are free of racing errors and tolerant to occasional collision of propagating SFQ pulses. This tolerance is due to stochastic nature of data signals generated by comparators operating within their gray zone. The circuits were fabricated in the eight-niobium-layer fabrication process SFQ5ee developed for superconductor electronics at MIT Lincoln Laboratory.

## I. Introduction

THIS paper is devoted to the development of the family of bioSFQ logic/memory cells proposed in [1] where we presented our basic ideas and the first experimental demonstrations. Here, we will present recent advances and more accurately define the ultimate goals and boundaries of the proposed bioSFQ technology.

For more than 30 years, superconductor single flux quantum (SFQ) electronics has been bounded by a widely accepted RSFQ approach [2], despite the existence of some applications for which RSFQ is not a particularly suitable technology. For instance, from the very beginning, it has been known that "a universal von-Neumann-type computer is probably the worst device for implementation using the RSFQ … technology" [2]. The technology has survived due to impressive progress in digital signal processing applications using fast ADCs and DACs; see, e.g., [3]−[5]. However, there was always a hope that RSFQ technology is just waiting for inherently matched applications. These days it is becoming more evident to us that neuromorphic computing may be such a long-awaited application. Our belief is supported by appearance of a number of recent proposals of neuromorphic devices containing RSFQ components.

Currently, artificial intelligence (AI) and, in particular, deep learning algorithms run on conventional purely digital computers. However, there is a general belief that special, not-yet-invented, computing devices could be much more efficient for AI applications. Carver Mead explained his belief in the following way: "For … problems, … in which the input data are ill-conditioned and the computation can be specified in a relative manner, biological solutions are many orders of magnitude more effective than those we have been able to implement using digital methods" [6]. He believed that: "Large-scale adaptive analog systems are more robust to component degradation and failure than are more conventional systems, and they use far less power" [6].

To adapt these recommendations, we supplemented purely digital RSFQ devices by a number of analog devices. We named the created combination as bioSFQ logic/memory family [1]. By creating the bioSFQ family we have built new bridges between the analog and digital domains of superconductor electronics. These bridges are Current to Pulse Rate (CPR) and Pulse Rate to Current (PRC) converters. CPR conversion is performed by a comparator, while PRC conversion is performed by an *LR* circuit. We have developed an analog bipolar multiplier, presented in Sec. II, to illustrate advantages of combining the analog and digital domains for neuromorphic

Manuscript receipt and acceptance dates will be inserted here. This work was supported by the Under Secretary of Defense for Research and Engineering under Air Force Contract No. FA8702-15-D-0001. *(Corresponding author: Sergey Tolpygo.)*
V. K. Semenov is with the Department of Physics and Astronomy, Stony Brook University, Stony Brook, NY 11794-3800, USA (e-mail: Vasili.Semenov@StonyBrook.edu).

E. B. Golden and S. K. Tolpygo are with the Lincoln Laboratory, Massachusetts Institute of Technology, Lexington, MA 02421, USA (e-mails: Evan.Golden@ll.mit.edu, Sergey.Tolpygo@ll.mit.edu). Color versions of one or more of the figures in this paper are available online at http://ieeexplore.ieee.org.

Digital Object Identifier will be inserted here upon acceptance.

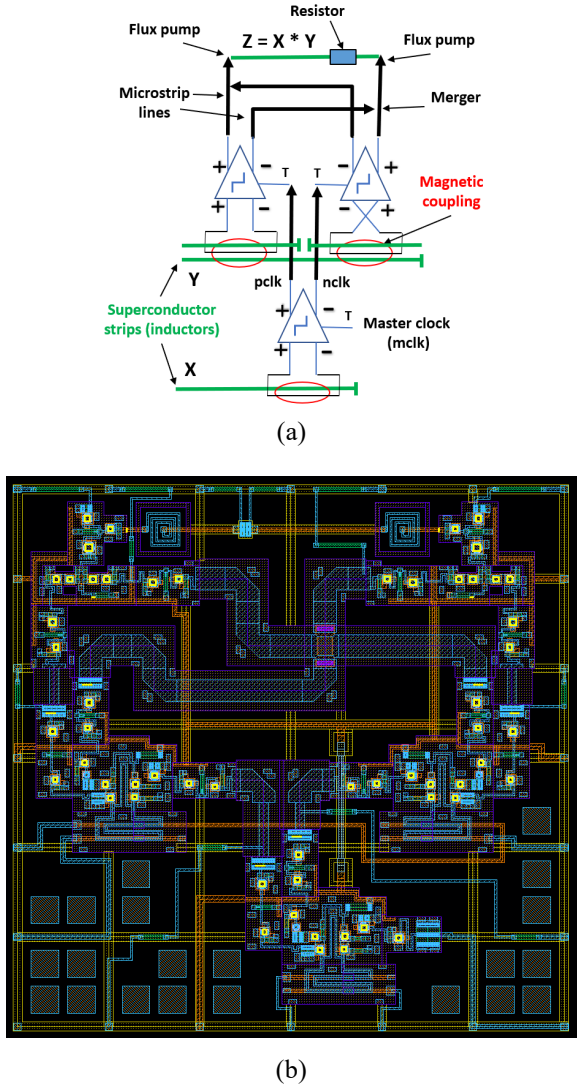

Fig. 1. (a) Block diagram of a bipolar analog multiplier using comparator with complementary outputs; (b) layout of the multiplier. The circuit size in (b) is 110 μm x 110 μm.

computing. We tend to think that the analog domain is preferable for a range of linear operations like summation, subtraction, differentiation, and integration, and for generating some non-linear functions, e.g., $\sin x$. In contrast, the digital domain is preferable for highly nonlinear logical operations and can be implemented in various superconductor digital technologies, in particular, by means of RSFQ circuitry.

## II. ANALOG MULTIPLIER

An analog multiplier was suggested in [1]. The block diagram of the present version of the multiplier is shown in Fig. 1. It has a few differences in wiring from the original block diagram presented in [1, Fig. 4a]. Moreover, some of its components, including the comparator with complementary outputs, have been significantly revised to make it simpler and more compact. As a result, the multiplier dimensions have been shrunk to 120 μm x 120 μm, reducing its area by a factor of 5 from the original design in [1].

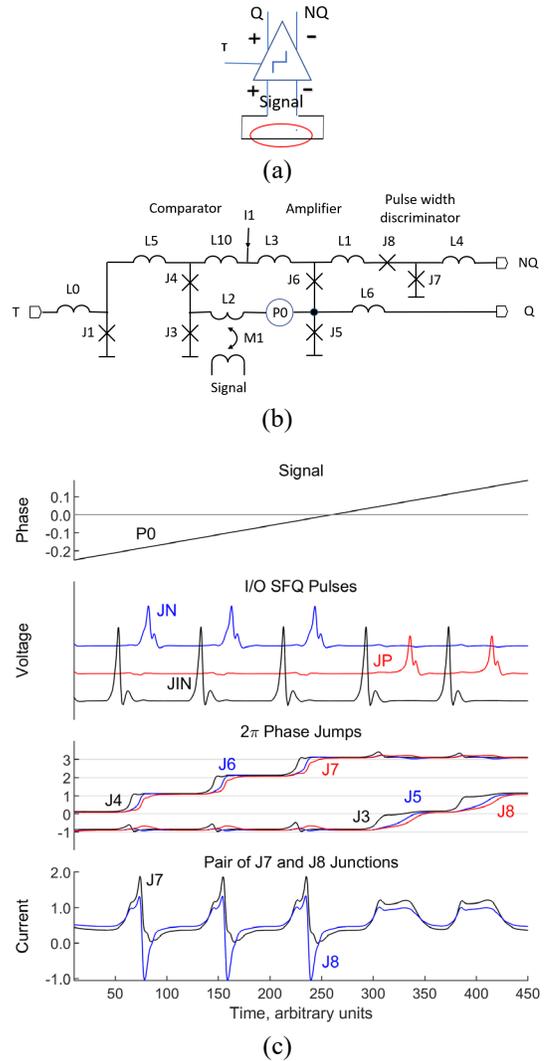

Fig. 2. (a) Symbolic notation; (b) equivalent circuit; (c) simulated operation of the comparator with complementary outputs. The simulation has been performed at the following set of parameters: I1=0.422 mA; L0=1.53 pH; L1=1.61 pH; L2=5.89 pH; L3=4.57 pH; L4=2.90 pH; L5=2.11 pH; L6=3.01 pH; L10=1.19 pH; J1=($I_c$ =0.279 mA, $\beta_c$ =0.25); J3=(0.14, 2.0); J4=(0.91, 2.0); J5=(0.196, 0.04); J6=(0.154, 2.0); J7=(0.09, 2.0); J8=(0.125, 0.01). The first number in parenthesis describing the properties of Josephson junctions is the critical current in mA, the second is the dimensionless $\beta_c$. The panels in (c) show from top to bottom: the input signal created by the phase source P0; five input SFQ pulses applied to the terminal **T** (trace JIN) and the "positive" (trace JP) and "negative" (trace JN) pulses produced at the Q and NQ output of the comparator, respectively; phase drops across the JJs; and currents in junctions J7 and J8 of the complementary output NQ. See text for the details.

### A. Comparator With Complementary Outputs

A nondestructive comparator described in [1, Fig. 1] is the key component of the unipolar bioSFQ cell family. In order to implement a more general bipolar mode of operation, in [1] we used an inverter cell along with a comparator to produce "positive" and "negative" trains of the SFQ pulses at the outputs of the composite circuit. Here we present a more compact and advanced solution − a comparator with the dual-rail output − allowing for a larger scale of integration. We preserved the same symbolic notation, shown in Fig. 2a, for the new circuit as was used for the previous version of the comparator in [1].

The new comparator shown in Fig. 2b consists of the following parts: a conventional comparator using Josephson junctions J1, J3 and J4; an amplifier (junctions J5 and J6); and a pulse width discriminator (junctions J7 and J8) built as a highly asymmetric comparator. An analog signal is applied via a magnetic coupling, M1, with inductor L2. For numerical simulations, the coupling transformer was represented by an ideal phase source, P0, in series with L2.

Correct operation of the comparator requires optimization of the dimensionless McCumber parameter $\beta_c$ of the individual junctions. This differs from the common practice of using Josephson junctions with identical $\beta_c$ values. The optimized $\beta_c$ along with the critical currents of the JJs are given as properties of the junctions in the format $(I_c, \beta_c)$ in the caption of Fig. 2. Parameters $\beta_c$ were adjusted by selecting the proper resistive shunts. Evidently this procedure changes values of the $I_c R_n$ products, i.e., the characteristic voltages $V_c$ of the junctions. We omitted numerical values of the characteristic voltages because they can be extracted from the given McCumber parameters $V_c = (\beta_c I_c \Phi_0 / 2\pi C_s)^{1/2}$, where $C_s = 70$ pF/μm² is the junction specific capacitance in the SFQ5ee process used [9], [10]. The smallest implemented $\beta_c=0.01$ corresponds to $V_c \approx 7$ μV. This does not affect the clock rate of the circuit, which was tested at clock rates up to ~37 GHz, due a special function of the corresponding junction, see below.

Results of the simulations using PSCSAN2 [11] are shown in Fig. 2c. The upper trace shows the phase difference created by the P0 representing the coupled input signal. It linearly grows from about −0.2 to 0.2 (in units of 2π). During the same time interval, five input SFQ pulses are applied to terminal **T**; they are shown by the black trace marked JIN in the second from the top panel in Fig. 2c. As expected, at the negative polarity of the input signal, the circuit responds by producing three SFQ pulses, marked JN, on the complementary terminal NQ. Alternatively, at the positive input signal, the circuit responds by producing two SFQ pulses, marked JP, on the terminal Q. The latter is the standard response for all traditional Josephson comparators [7], [8], while the former has been enabled by the development of the complementary, NQ, output in this research. The presented voltage traces JIN, JN and JP show voltage drops on external junctions attached to the input and output terminals and not shown in Fig. 2b.

The traces shown in the panel marked "2π Phase Jumps" in Fig. 2c, illustrate dynamics of the internal Josephson junctions. Crudely, it could be described as two independent sequences of 2π-phase leaps on some of the Josephson junctions. At a negative value of the input signal, the upper junction, J4, of the comparator J3−J4 switches. The SFQ pulse propagates via junctions J6 and J7 to the output terminal NQ. At a positive input signal, the lower junction, J3, of the comparator converts this signal to the SFQ pulse and junction J5 passes this pulse to the terminal Q. Junction J8 prevents propagation of the "positive" pulses to the terminal NQ. In turn, junction J5 prevents propagation of the "negative" pulses to the terminal Q.

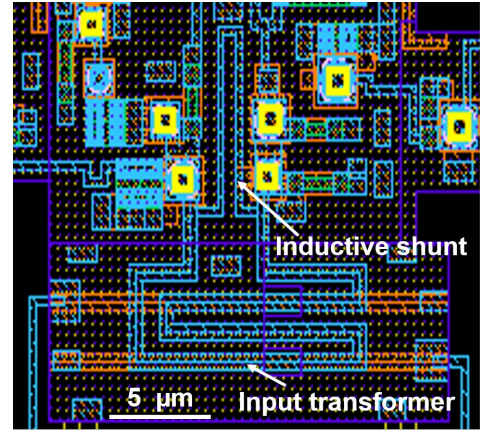

Fig. 3. Fragment of Fig. 1b, showing the input transformer – the blue meander. Inductive shunt is used to change the transformer coupling efficiency and the effective gray zone width of the comparators.

The described "isolation" technique is rarely if ever used in SFQ circuits, though a similar idea was used for implementing a NOT element in the original resistive SFQ [12]. At first glance, it is difficult to imagine how junction J8 distinguishes or discriminates SFQ pulses generated by junctions J5 and J6. Such discrimination is possible if, for example, junctions J5 and J8 have much lower $\beta_c$ than junction J7, and the critical current of J8 is lower than that of J7. In this case, a slow-responding J8 is insensitive to short SFQ pulses generated by J6. At the same time, it easily responds to much wider pulses generated by J5.

Two bottom traces in Fig. 2c show currents via J7 and J8, and illustrate the presented isolation technique. During the negative signal polarity (from time zero to 275 a.u.), the short pulses generated by J6 flow through J8 and activate J7, whereas wide pulses generated by J5 (from time 275 to 450 a.u.) slowly increase the bias of J8 over its critical value and hold it until the junction phase completes its 2π leap.

### B. Input Transformer of the Comparator

As mentioned in [1], we need different comparators for copying analog currents and for more complex arithmetic operations with such currents. In the first case, the gray zone of the comparator should be as narrow as possible. This is a conventional requirement, and has been addressed in a number of publications, see, for example [7], [8]. In the second case, the gray zone should be stretched to the whole range of the input current variations. To achieve this goal, we shunt the transformer, shown in Fig. 3 by a blue meandering trace overlapping the orange horizontal traces, by an inductor seen in Fig. 3 as a blue vertical ⊓-shape.

### III. FABRICATION AND MEASUREMENTS

The comparator and multiplier circuits were fabricated in the eight-niobium-layer fabrication process SFQ5ee [9] with the Josephson critical current density of 100 μA/μm² and minimum linewidth of 0.25 μm, developed for superconductor electronics at MIT Lincoln Laboratory. The circuits were tested in a low-speed liquid helium dunk probe using a multichannel test setup Octopux [13].

*A. Comparator Testing*

The basic functionality of a single comparator with complementary outputs is shown in Fig. 4. A stream of SFQ pulses with approximately 31 GHz rate, $f_{cl}$, corresponding to a time-averaged input voltage $V_{in} = f_{cl}\Phi_o$ of approximately 64 µV, was generated on-chip at the input **T** in Fig. 2b. The stream of pulses divides between the two outputs, Q and NQ, according to the transfer function of the comparator and the sign and amplitude of the input (control) current, $I$, coupled into the inductor L2. The transfer function describes the probabilities of pulse passage $p_Q$ and $p_{NQ}$ within the gray zone of the comparator formed by J3-J4:

$$p_Q = \tfrac{1}{2}\left[1 + \mathrm{erf}\left(\pi^{1/2}\tfrac{I-I_{th}}{\Delta I}\right)\right] \text{ and } p_{NQ} = 1 - p_Q, \quad (1a)$$

where erf($x$) is the error function, $I_{th}$ is the adjustable threshold current of the comparator, and $\Delta I$ is the gray zone width modified by coupling constant of the input signal transformer M1.

The passing pulses generate the time-averaged voltage at each of the respective outputs

$$V_Q = p_Q V_{in} \text{ and } V_{NQ} = (1 - p_Q)V_{in}, \quad (1b)$$

which was measured and shown in Fig. 4 along with fits to (1b).

In the range of currents near the threshold $|I - I_{th}| < \Delta I$, expansion of the error function gives linear relations between the output voltages and the control current

$$V_Q = (\tfrac{1}{2} + \tfrac{I-I_{th}}{\Delta I})V_{in}; \ V_{NQ} = (\tfrac{1}{2} - \tfrac{I-I_{th}}{\Delta I})V_{in}. \quad (1c)$$

These linear relations are clearly visible in Fig. 4. This linear regime of the comparator operation is important for the use in bioSFQ circuits considered below: multipliers, dividers, etc.

The threshold current of the comparator was adjusted to be zero. The device was fully operational in a wide range of input frequencies from zero to about 36 GHz. Between about 36 GHz and 70 GHz, the gray zone broadened but preserved the linear relations (1c). At higher frequencies, significant deviations of the transfer function from (1) appeared. These deviations are similar to the ones observed on the traditional Josephson balanced comparators with the Q-output; see [14] for the most recent publication and references therein.

To summarize, we demonstrated that the comparator works as a bipolar current-to-SFQ pulse rate converter, CPRC.

*B. Testing Bipolar Multiplier*

The analog bipolar multiplier is shown in Fig. 1 and composed of three comparator-based circuits. For testing, the on-chip master clock with the SFQ pulse rate of about 30 GHz was applied to the terminal **T** of the bottom comparator, having the control current $X$, in Fig. 1a. The comparator converts the control current $X$ into two streams of pulses, Q and NQ, with frequencies $p_Q(X)f_{cl}$ and $p_{NQ}(X)f_{cl}$ depending on the value of $X$. Each stream serves as the clock for one of the two comparators controlled by the same current $Y$, see Fig. 1a. These two comparators differ only by the sign of the mutual inductance M1 of the input transformer (i.e., by the direction of winding) which couples the input current $Y$; see circled areas marked "Magnetic coupling" in Fig. 1a.

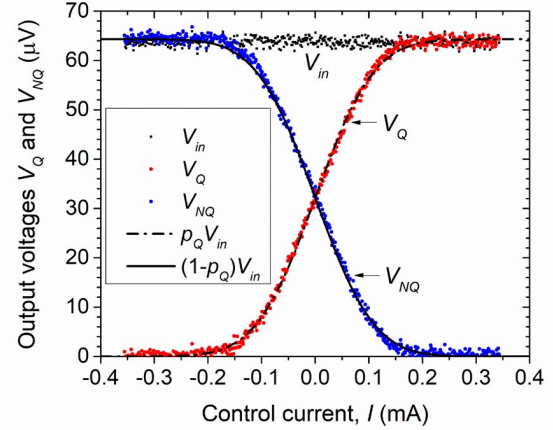

Fig. 4. Measurements of the comparator with complementary outputs, shown in Fig. 2b. The comparator output voltage on Q output, $V_Q$ (red dots) and NQ output, $V_{NQ}$ (blue dots) as a function of control current in the signal transformer primary; see Fig. 2b. The clock frequency $f_{cl}$ at terminal **T** was about 31 GHz, corresponding to the time-averaged voltage shown by the black squares and marked $V_{in}$. The solid black curve and the dash-dot black curve are the fits to (1b) with $p_Q$ given by the error function (1a). The fit parameters are $\Delta I = 0.218$ mA, $I_{th} = 0$, and $V_{in} = 64.3$ µV. Note a slight deviation of the $V_{NQ}$ data from (1) at high output voltages and large negative control currents, $I - I_{th} < -\Delta I$ (at high SFQ pulse rates) and a slight difference between $p_{NQ}(-I)$ and $p_Q(I)$. These small deviations occur at high input clock rates and are much smaller at $f_{cl}$ = 20 GHz. These distortions are not important for the devices operating in the linear regime (1c).

Each of the $Y$-controlled comparators generates streams of Q and NQ pulses with frequencies, respectively, $p_Q(Y)p_Q(X)f_{cl}$ and $p_{NQ}(Y)p_{NQ}(X)f_{cl}$, depending on the values of $X$ and $Y$. These streams are merged such that the Q stream of the left comparator (with the positive M1) is merged with the NQ stream of the negative fluxons from the right comparator (with the negative M1) and vice versa, as shown in Fig. 1a. The merging performs addition of the streams, creating two new streams of fluxons:

$$p_Q(Y)p_Q(X)f_{cl}\Phi_0 + p_{NQ}(-Y)p_{NQ}(X)f_{cl}(-\Phi_0) \quad (2a)$$

$$\text{and } p_{NQ}(Y)p_Q(X)f_{cl}(-\Phi_0) + p_Q(-Y)p_{NQ}(X)f_{cl}\Phi_0, \quad (2b)$$

where the $-Y$ is due to the negative sign of the mutual inductance in input transformer of the comparator on the right side of Fig. 1a.

The stream (2a) pumps positive flux and the stream (2b) pumps negative flux into an $LR$ circuit − the superconducting loop $L$ interrupted by resistor $R$; see the very top of Fig. 1a.

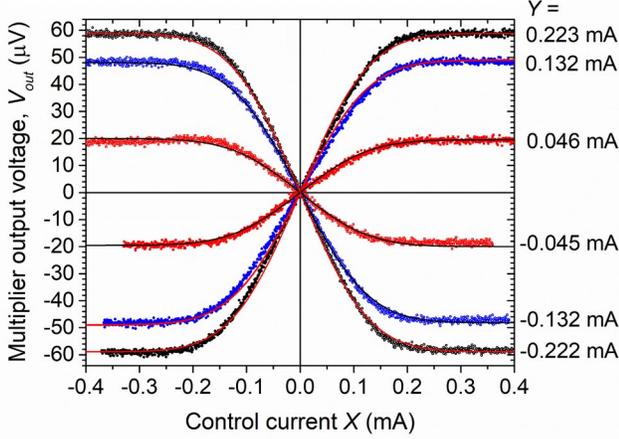

Fig. 5. Output voltage of the analog multiplier as a function of the control current $X$ at a few different values of the control current $Y$. The values of $Y$ are given on the right panel near each curve. The solid curves show the fits to (7). The $f_{cl}$ was about 30 GHz, corresponding to the time-averaged input voltage of about 60.5 µV. The fitting curves correspond to the following parameters of the comparators in (5): $V_{in}$ =60.54 µV, $\Delta X = \Delta Y = 0.253$ mA, and values $Y = 0.223, 0.132, 0.042, -0.043, -0.128, -0.222$ mA from the top to bottom, respectively. The latter values are very close to the measured control current values. The threshold currents of the comparators in the multiplier were adjusted to zero.

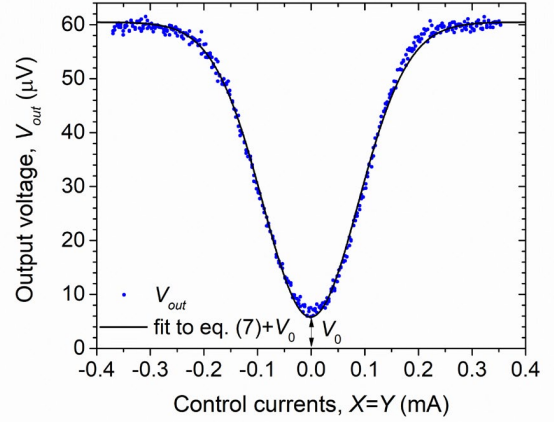

Fig. 6. Implementation of an "$X^2$" function $V_{out} = [erf(\pi^{1/2} X/\Delta)]^2 V_{in}$, using the bipolar multiplier with equal input currents $X=Y$. The threshold currents of the comparators were adjusted to zero values. Solid black curve shows a fit to (7) plus an offset $V_0$. The fitting gives $V_{in}$ =54.7 µV, $\Delta X = \Delta Y$ =0.253 mA, the same values as in Fig. 5, and $V_0$ =5.8 µV. A nonzero $V_{out}$ at $X = 0$ is likely caused by the imperfect adjustment of the threshold currents of the comparators and the test setup voltage offset. A small asymmetry, visible at large output voltages, apparently reflects performance of the real comparators at large control currents $|I - I_{th}| > \Delta I$; see Fig. 4.

These two streams of fluxons induce the total flux in the loop $L$ and set the current in the resistor $R$.

In the experiment, we measure the time-averaged voltage, $V_{out}$, across the resistor $R$ in Fig. 1a. This voltage is proportional to the current through the resistor and equals to the rate of change of the total flux in the loop

$$V_{out} = [p_Q(Y)p_Q(X) - p_{NQ}(-Y)p_{NQ}(X)]f_{cl}\Phi_0 - [-p_{NQ}(Y)p_{NQ}(X) + p_Q(-Y)p_{NQ}(X)]f_{cl}\Phi_0 . \quad (3)$$

For completely symmetrical comparators

$$p_Q(-Y) = p_{nQ}(Y) \text{ and } p_{NQ}(-Y) = p_Q(Y), \quad (4)$$

which follow from (1); see also Fig. 4. Then (3) reduces to

$$V_{out} = [p_Q(Y) - p_{NQ}(Y)][p_Q(X) - p_{NQ}(X)]V_{in} =$$
$$= V_{in} \, \text{erf}\left(\pi^{1/2} \frac{Y}{\Delta Y}\right) \text{erf}\left(\pi^{1/2} \frac{X}{\Delta X}\right), \quad (5)$$

where $\Delta X, \Delta Y$ are the widths of the gray zone of the $X$-controlled and $Y$-controlled comparators. So, the circuit performs an analog multiplication of two error functions of the control currents $X$ and $Y$, and the resulting product is encoded as the time-averaged output voltage, $V_{out}$.

The test results showing the described operation of the analog bipolar multiplier are given in Fig. 5. A few different values of $Y$ were supplied to the circuit, corresponding to differently colored traces in Fig. 5, while the value of $X$ was continuously swept. The threshold currents $I_{th}$ of the comparators were adjusted to zero. At small values of the control currents $X \ll \Delta X, Y \ll \Delta Y$, the circuit performs a simple multiplication $XY/(\Delta X\Delta Y)$, as follows from the expansion of the error function

$$V_{out} = \frac{XY}{\Delta X \Delta Y} V_{in} . \quad (6)$$

For the identical comparators, their gray zone widths are equal, $\Delta X = \Delta Y \equiv \Delta$.

At $X = Y$, the multiplier output voltage is

$$V_{out} = V_{in}[\text{erf}\left(\pi^{1/2} \frac{X}{\Delta X}\right)]^2. \quad (7)$$

In the linear regime (at small control currents), the circuit performs $X^2/\Delta^2$ operation.

Testing the multiplier operation (7) at $X = Y$ is shown in Fig. 6. The required synchronization of the control currents can be achieved by connecting the control lines in series. The pure $X^2$ function can be seen in Fig. 6 at small values of the control currents. A small offset $V_0$ is likely an artifact of the measurements.

## IV. DIVIDER AND OTHER BIOSFQ CIRCUITS

The demonstrated analog multiplier is the key building block for several other analog arithmetic devices. In the previous section we showed how to use the multiplier to execute the $X^2$ function. In [1], we explained how a sensitive comparator can be used in a feedback loop to enable copying of currents, i.e., generating a current $Z$ equal to the control current $X$; see [1, Fig. 3]. Using the developed multiplier and this Copy function, we can build an analog divider performing Y/X function on two analog currents, i.e., generating a current Z equal the ratio of two input currents: Z = Y/X. The block diagram of the bipolar divider circuit is shown in Fig. 7. The divider is composed of the multiplier, very similar to the one shown in Fig.1a, and the comparator shown in Fig. 2. The comparator compares current Y and the result of multiplication. The mismatch of the comparison is converted by the comparator into a difference of streams of the positive and negative SFQ pulses applied to strip (inductor) Z, the second input of the multiplier. When the mismatch is eliminated and the result of multiplication equals the Y, the current Z (in the stirp Z) equals Y/X. This is because Z·X = Y.

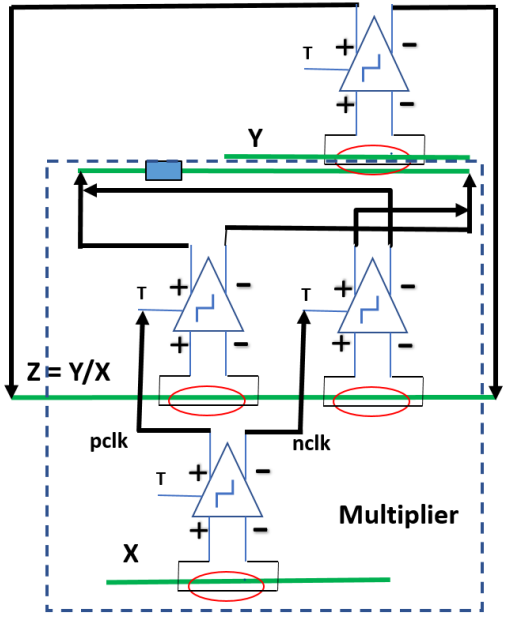

Fig. 7. Block diagram of a bioSFQ divider circuit performing $Y/X$ operation on two currents $Y$ and $X$.

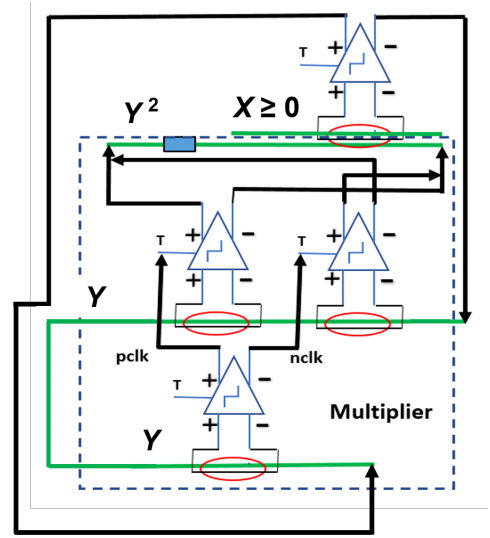

Fig. 8. Block diagram of a bioSFQ circuit performing analog square root operation on a positive input current $X$, $Y = \sqrt{X}$; see text.

As discussed in [1], the "comparing" comparator (at the very top of the Fig. 7) should have a narrow gray zone, contrary to the comparators in the multiplier which should have sufficiently wide gray zones to accommodate the full expected range of input currents Y and X.

Using a similar approach, we can design a bioSFQ circuit performing a square root operation, i.e., a circuit generating a current $Y = X^{1/2}$ for $X \geq 0$. The block diagram of the circuit is shown in Fig. 9. The circuit uses a Copy function from [1] to compare the input current X with the result of the $Y^2$ operation of the multiplier. The multiplier slightly differs in wiring from the one shown in Fig. 1a. The feedback SFQ pulses generated at the Q and NQ outputs of the Copy comparator, at the very top of Fig. 8, are converted into the Y current applied in series to both inputs of the multiplier. After the difference $Y^2 - X$ is eliminated by the feedback loop, the Y current will be equal to $\sqrt{X}$.

Other simple functions can also be executed using the proposed approach.

## V. DISCUSSION AND FUTURE WORK

Our ultimate goal is to develop a number of analog and hybrid analog/digital computing and controlling circuits for various applications in neuromorphic computing, controlling processors, e.g., for quantum annealers and gate-based quantum computing, etc.

The simplified structure of an artificial superconductor neuron is shown in Fig. 9a. Multiple input SFQ streams merge into one stream that induces an electric current, $I$, in a loop with large inductance $L$. A Josephson junction, $J$, serves as a fuse and allows magnetic flux quanta to exit the loop via the SFQ Output. Without resistor $R$ in series with $L$, the intensity of the output stream would equal the sum of intensities of the inputs. However, a fraction of magnetic flux escapes the loop via the resistor. Consequently, the induced current $I$ decays with characteristic time $\tau = L/R$. As a result, the "fuse" junction $J$ is opened only at intensities of the input streams exceeding some predefined value set by the junction critical current. So, the described circuit behaves similarly to the expected behavior of a conventional artificial neuron.

Figure 9b shows a structure of a more advanced bipolar neuron that unites functionalities of conventional excitatory and inhibitory neurons [15]. The flux quanta coming from the excitatory inputs increase current $I$, while the flux quanta from inhibitory inputs decrease it. The comparator applies the result of the comparison of the current $I$ and the threshold current $Ith$ to the complementary (excitatory and inhibitory) outputs. We note that the bipolar version of the superconductor neuron can be recognized as a part of the multipliers described in Sec. II-IV; see Figs. 1a, 7, 8.

In the framework of the deep learning paradigm, explained, e.g., in [16], neurons are organized into connected layers as shown in Fig 10a. In particular, the recognition of, say, an input optical image, could be described as a propagation of an "active processing zone" by a feedforward way, i.e., from left to right in Fig. 10a. The learning process is required to reduce the recognition errors. A backpropagation of error ('backprop') is the most often used algorithm to train deep neural networks and is the most successful learning procedure for these networks [17]. In the framework of this algorithm, the active zone, where corrections are calculated, counterflows to the recognition process. Schematically the backpropagation algorithm is mapped in Fig. 10b.

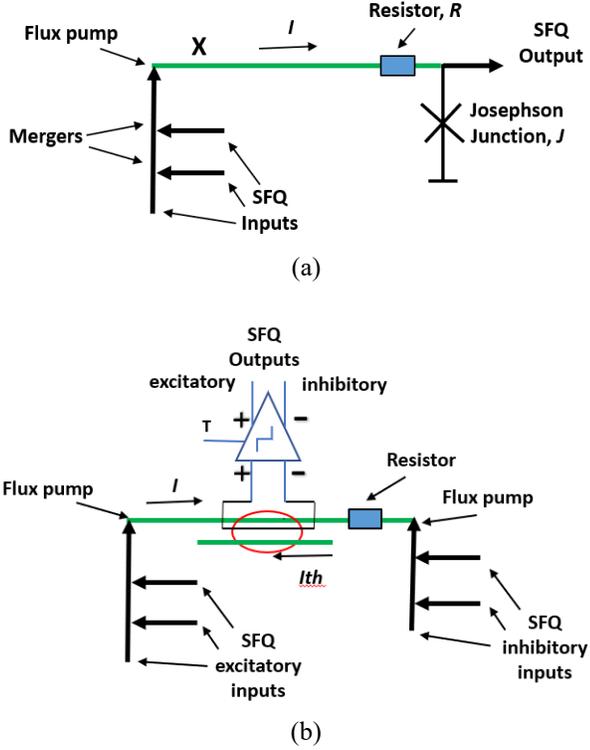

Fig. 9. Structures of superconductor artificial neurons: (a) unipolar neuron; (b) bipolar neuron. Green lines correspond to superconducting inductors with large inductance values.

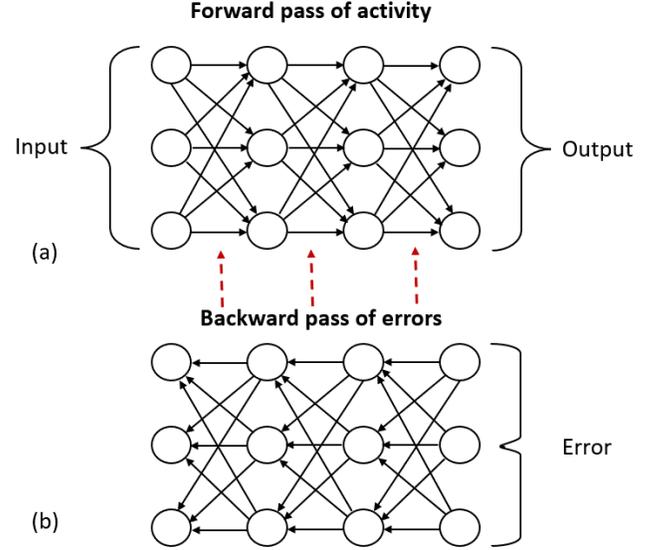

Fig. 10. Backpropagation algorithm.

Using a conventional computer, it is easy to compute and store memories of separate feedforward and feedback passes in order to take differences and then use them for learning [16]. We believe that bioSFQ cell family will allow us to implement the backprop algorithm in a more neuromorphic way. The prospective bioSFQ solution is exhibited in Fig. 10 by the three red arrows. Our current effort is on designing a scalable array of bioSFQ circuits similar to the arrays shown in Fig. 9 with "comparing" comparators providing the required backpropagation feedback. The desired functionality could be reprogrammed by rerouting feedback streams of SFQ pulses. Several rerouting techniques have been suggested; see, e.g., [18].

## VI. Conclusion

We have demonstrated a comparator with complementary outputs and an analog bipolar multiplier for the use in bioSFQ circuit family, proposed in [1], for neuromorphic and hybrid analog/digital computing. We also gave a few examples of neuromorphic circuits using the analog bipolar multiplier demonstrated in this work. The results show that we are quite close to implementing a wide range of mixed analog/digital functions, using the proposed circuitry.

It can be noted that novel neuromorphic circuits are heavily supported by the conventional digital circuitry; see, e.g., [19]. Similarly, it is safe to assume that superconductor neuromorphic circuits will also be heavily supported by superconductor digital electronics. In all the demonstrated bioSFQ circuits we used RSFQ [2] components and the RSFQ dc biasing scheme because they are currently the most developed and, therefore, the most suited candidates. However, for reducing energy dissipation and the dc bias currents, all modern innovations such as energy efficient biasing [20], bias current recycling [21], [22], ac/dc conversion [23], and SFQ biasing [24] can be readily implemented without any changes to the basic bioSFQ approach. Also, there are no reasons why other digital technologies such as RQL [25] or (adiabatic) quantum flux parametron [26], [27] cannot be implemented in a similar manner in the described or similar superconductor neuromorphic circuits.

A path toward decreasing the energy dissipation and increasing the scale of integration would be decreasing the junction sizes and critical currents as well as implementing self-shunted Josephson junctions. According to Mead [6], neuromorphic applications are much more tolerant to error rate than conventional digital applications. We think that 0.1% and in some cases 1% error rate could be quite satisfactory. In practical terms, it means that critical currents of junctions in bioSFQ cells could be safely reduced from 0.1 mA to, say, 0.02 mA. This task could be easily executed. However, scaling down the critical currents requires inversely proportional increase of all cell inductances. This task is quite challenging in the fabrication processes using Nb wiring because this requires a greatly increased length of the inductors. Fortunately, a novel fabrication technology using bilayers of kinetic and geometrical inductors, NbN/Nb bilayers, has been developed [28], [29] and enables compact kinetic inductors at the same time preserving efficiency of the required inductive couplers and transformers.

ACKNOWLEDGMENT

The numerical simulations were performed using PSCAN2 software package developed by Pavel Shevchenko [11]. We thank to Coenrad Fourie for assistance with InductEx software [30] used for inductance extraction from circuit layouts. We are also grateful to Vlad Bolkhovsky and Ravi Rastogi for overseeing the wafer fabrication.

This research was supported by the Under Secretary of Defense for Research and Engineering under Air Force Contract No. FA8702-15-D-0001. Any opinions, findings, conclusions or recommendations expressed in this material are those of the author(s) and do not necessarily reflect the views of the Under Secretary of Defense for Research and Engineering. Delivered to the U.S. Government with Unlimited Rights, as defined in DFARS Part 252.227-7013 or 7014 (Feb 2014). Notwithstanding any copyright notice, U.S. Government 11 rights in this work are defined by DFARS 252.227-7013 or DFARS 252.227-7014 as detailed above.